\documentclass[sigconf]{acmart}
\AtBeginDocument{%
  }

\usepackage{multirow}
\usepackage{graphicx}
\usepackage{subcaption}
\usepackage[percent]{overpic}
\setcopyright{acmlicensed}
\copyrightyear{2026}
\acmYear{2026}
\acmDOI{XXXXXXX.XXXXXXX}
\acmConference[]{}{}

\begin{document}

\title{ST-FateFlow: Continuous Gene-Spatial Reconstruction with Differentiation-Potential Learning from Time-Resolved Spatial Transcriptomics}

\author{Zhe Yu}
\affiliation{%
  \institution{State Key Laboratory for Novel Software Technology, Nanjing University}
  \city{Nanjing, Jiangsu}
  \country{China}}
\affiliation{%
  \institution{Shenzhen Institute of Advanced Technology, Chinese Academy of Sciences}
  \city{Shenzhen, Guangdong}
  \country{China}}
\affiliation{%
  \institution{Faculty of Computer Science and Control Engineering, Shenzhen University of Advanced Technology}
  \city{Shenzhen, Guangdong}
  \country{China}}
\email{yuz@smail.nju.edu.cn}

\author{Xiaobo Sun}
\authornotemark[1]
\affiliation{%
    \institution{Department of Human Genetics, Emory University}
    \city{Atlanta, GA}
    \country{USA}
}
\email{xiaobo.sun@emory.edu}

\author{Hao Wu}
\authornote{corresponding authors}
\affiliation{%
    \institution{Shenzhen Institute of Advanced     Technology, Chinese Academy of Sciences}
    \city{Shenzhen, Guangdong}
    \country{China}
}
\affiliation{%
  \institution{Faculty of Computer Science and Control Engineering, Shenzhen University of Advanced Technology}
  \city{Shenzhen, Guangdong}
  \country{China}}
\email{wuhao@suat-sz.edu.cn}

\renewcommand{\shortauthors}{Zhe Yu, Xiaobo Sun and Hao Wu.}

\begin{abstract}
  Time-resolved spatial transcriptomics provides a powerful view of dynamic biological processes by measuring cellular states together with their spatial organization across time. However, destructive profiling and experimental cost restrict observations to a few discrete time points collected from independent samples, with no known cell pairing across time.  This makes it difficult to reconstruct gene expression and tissue organization at unobserved stages and to characterize differentiation-associated dynamics. Here, we introduce ST-FateFlow, a flow-matching framework for continuous gene-spatial dynamics and differentiation-potential learning. ST-FateFlow jointly learns gene-expression and spatial-coordinate velocity fields, enabling interpolation of missing time points and continuous modeling of transcriptional evolution and tissue remodeling. In parallel, it independently learns a time-conditioned differentiation potential, whose values organize cell states along inferred differentiation progressions and support the analysis of differentiation-associated transcriptional changes. Across multiple time-resolved spatial transcriptomics datasets, ST-FateFlow effectively captures coordinated gene-expression and spatial dynamics. Case studies further show that the learned potential landscapes are consistent with known developmental and regenerative trajectories and help identify candidate genes associated with differentiation progression.
\end{abstract}



\maketitle
\raggedbottom

\section{Introduction}
Development and regeneration involve coordinated changes in cellular molecular states and tissue organization over time \cite{gerri2020human, liu2024spatiotemporal, lazar2026spatial}. Time-resolved spatial
transcriptomics provides snapshots of these processes by jointly measuring gene expression and spatial coordinates at multiple stages \cite{staahl2016visualization, rodriques2019slide, wei2022single, oliveira2024characterization}. However, destructive profiling prevents the same cells from being followed over time, while experimental cost usually limits observations to a small number of independently sampled tissues. The continuous molecular transitions and tissue remodeling between observed stages therefore remain unmeasured.

Recovering these missing dynamics would provide a continuous view of how transcriptional changes accompany tissue reorganization. Beyond predicting intermediate states, an interpretable potential landscape can help organize cell populations along differentiation progressions and characterize transcriptional changes associated with transitions from progenitor-like to differentiated states. These objectives are complementary but impose different modeling requirements. Accurate reconstruction benefits from a flexible vector field capable of describing complex population transport, whereas a dynamics-consistent potential landscape can be obtained by constraining an auxiliary field to be the negative gradient of a potential function.

Existing methods address different parts of this problem. Temporal models such as Waddington-OT \cite{schiebinger2019optimal}, scNODE \cite{zhang2024scnode}, and PRESCIENT \cite{yeo2021generative} infer population dynamics primarily in gene-expression space without jointly reconstructing tissue geometry. Spatial alignment and spatiotemporal modeling approaches, including DeST-OT \cite{halmos2025dest}, MOSCOT \cite{klein2023mapping}, Spateo \cite{qiu2024spatiotemporal}, and stVCR \cite{peng2026stvcr}, align tissue sections or model continuous gene-spatial changes, but do not provide a directly evaluable differentiation potential. Conversely, STORIES learns differentiation trajectories and potential landscapes from spatiotemporal transcriptomic data, but is not designed to generate continuous joint gene-expression and spatial-coordinate trajectories at unobserved stages \cite{huizing2026stories}.

These considerations motivate a framework that separates continuous reconstruction from differentiation-state organization while grounding both in the same inferred cross-time transitions. The reconstruction objective requires a flexible gene-spatial velocity field that can capture complex transcriptional changes and tissue deformation. However, such local and high-dimensional velocities do not directly indicate where an individual cell lies along the overall differentiation progression. We therefore introduce a parallel potential branch whose negative gradient is trained to follow the same earlier-to-later transition directions inferred from adjacent snapshots. Consequently, the learned potential is encouraged to decrease locally along data-supported transitions, allowing it to organize cell states along inferred differentiation progressions rather than represent a physical energy.

To implement this idea, we propose ST-FateFlow, a flow matching framework \cite{lipman2022flow, tong2023conditional} for continuous gene-spatial reconstruction and differentiation-potential learning from sparse, unpaired spatial transcriptomic snapshots. ST-FateFlow first uses fused Gromov–Wasserstein unbalanced optimal transport to infer a structure-aware transport coupling between adjacent time points and performs coupling-guided rigid registration to remove global coordinate discrepancies. It then learns a gene-spatial velocity field using local flow-matching supervision together with coordinate distribution matching for global tissue-geometry preservation. In parallel, a separately parameterized potential function is trained from the same coupling-derived transition velocities, providing directly evaluable potential values for comparing cell states along inferred differentiation progressions and analyzing associated gene-expression changes. Our experiments across developmental and regenerative datasets show that ST-FateFlow accurately reconstructs molecular and spatial dynamics, reproduces differentiation orderings consistent with established trajectories, and identifies genes associated with regenerative-to-neuronal state transitions.

Our main contributions are:
\begin{itemize}
\item We develop a continuous gene-spatial modeling framework that combines structure-aware unbalanced transport, rigid registration, and flow matching to reconstruct molecular and spatial dynamics from unpaired time-resolved spatial transcriptomic snapshots.
\item We introduce an independently trained potential branch whose negative state gradient is aligned with inferred earlier-to-later transitions. This branch provides a time-conditioned potential for comparing cell states along inferred differentiation progressions, while the gene-spatial velocity field remains flexible and is not required to be gradient-based.
\item Experiments across developmental and regenerative datasets demonstrate accurate interpolation of gene-expression distributions and tissue geometry, differentiation orderings consistent with established trajectories, and identification of genes associated with regenerative-to-neuronal state transitions.
\end{itemize}

\section{Related Work}

Existing methods related to our study can be broadly grouped into temporal single-cell modeling, spatiotemporal transcriptomics modeling, and cell-fate landscape learning.

\textbf{Temporal single-cell dynamics.}
RNA velocity estimates the direction of cellular state transitions from spliced and unspliced RNA measurements, while scVelo extends this idea using a dynamical model of transcriptional kinetics \cite{la2018rna, bergen2020generalizing}. When cells are collected at multiple experimental time points, Waddington-OT \cite{schiebinger2019optimal} infers population-level couplings between adjacent snapshots using optimal transport. PRESCIENT \cite{yeo2021generative} learns potential-driven cellular dynamics from time-series scRNA-seq data, and scNODE \cite{zhang2024scnode} uses neural ordinary differential equations to reconstruct continuous gene-expression distributions from discrete observations. These methods establish important foundations for temporal reconstruction but operate primarily in gene-expression space without modeling spatial tissue dynamics.

\textbf{Spatiotemporal transcriptomics modeling.}
Time-resolved spatial transcriptomic analysis first requires relating observations collected from different tissue sections. DeST-OT \cite{halmos2025dest} and MOSCOT \cite{klein2023mapping} use optimal transport to align spatial transcriptomic datasets across time. Moving beyond discrete alignment, Spateo \cite{qiu2024spatiotemporal} learns vector fields that describe spatial tissue deformation, whereas stVCR \cite{peng2026stvcr} jointly models changes in gene expression, spatial coordinates, and cell populations. These methods enable spatial alignment or continuous spatiotemporal reconstruction but do not provide an explicit differentiation potential for cell-state analysis.

\textbf{Cell-fate landscape learning.}
CellRank combines RNA velocity with Markov-state modeling to estimate fate probabilities and terminal states \cite{lange2022cellrank}. In spatial transcriptomics, SpaTrack reconstructs spatially informed differentiation trajectories and identifies genes varying along these trajectories \cite{shen2025inferring}. STORIES \cite{huizing2026stories} further learns differentiation trajectories and a potential landscape from time-resolved spatial transcriptomic data. These approaches provide interpretable descriptions of differentiation progression but do not simultaneously reconstruct continuous gene-expression and spatial-coordinate dynamics.

In summary, existing methods provide complementary capabilities for temporal trajectory inference, spatial alignment, continuous spatiotemporal reconstruction, and cell-fate landscape analysis. However, these methods typically address continuous gene-spatial reconstruction and differentiation-potential analysis separately. ST-FateFlow bridges these capabilities by learning a flexible gene-spatial velocity field and an independently parameterized time-conditioned potential from shared structure-aware cross-time transitions. The former supports continuous interpolation of molecular states and tissue geometry, while the latter organizes cell states along inferred differentiation progressions.

\begin{figure*}[t]
    \centering
    \includegraphics[width=0.98\textwidth]
    {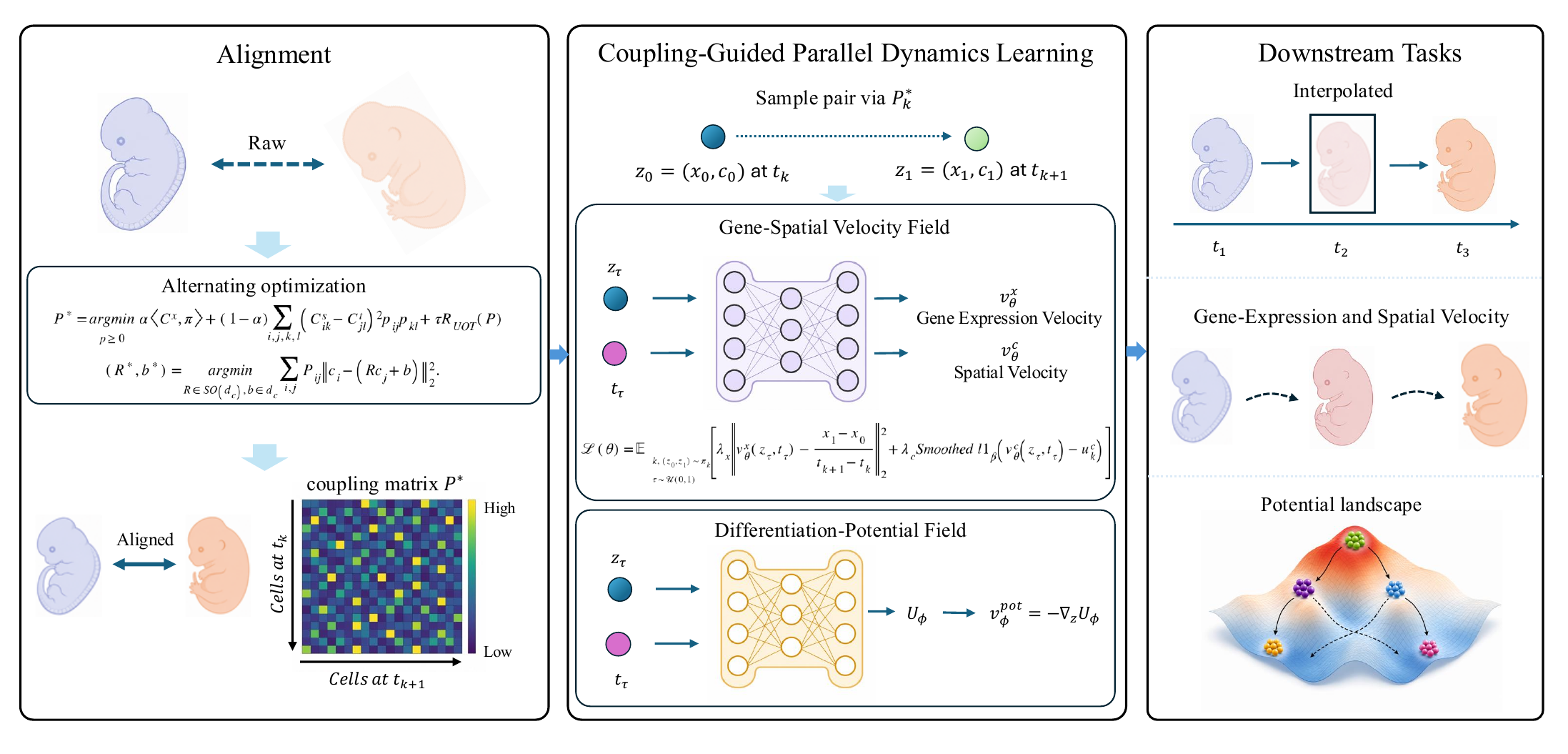}
    \caption{
    Overview of ST-FateFlow
    }
\end{figure*}

\section{Method}

\subsection{Problem Formulation}

Consider a sequence of spatial transcriptomic snapshots observed at time points
$\{t_k\}_{k=0}^{K}$. Snapshot $k$ contains $n_k$ cells and is represented as

\begin{equation}
\mathcal{S}_k=\left\{z_i^k=\left(x_i^k,c_i^k\right)\right\}_{i=1}^{n_k},\qquad
z_i^k\in\mathbb{R}^{d_x+d_c},
\end{equation}

where $x_i^k\in\mathbb{R}^{d_x}$ denotes the gene-expression profile of cell $i$, and $c_i^k\in\mathbb{R}^{d_c}$ denotes its spatial coordinate. The spatial dimension is denoted by $d_c\in\{2,3\}$, allowing the framework to
model both two-dimensional tissue sections and three-dimensional tissue reconstructions. The spatial dimension remains fixed across all snapshots within the same dataset.

Because spatial transcriptomic profiling is destructive, cells observed at different time points are unpaired. Moreover, different tissue samples may exhibit global rotations, translations, and changes in cell abundance. Our objective is to infer coupling between adjacent snapshots and learn a continuous joint velocity field:

\begin{equation}
v_\theta(z,t)=\left(v_\theta^x(z,t),v_\theta^c(z,t)\right),
\end{equation}

where $v_\theta^x$ describes gene-expression dynamics and $v_\theta^c$ describes spatial-coordinate dynamics. Given an initial cell state, its continuous trajectory is defined by

\begin{equation}
\frac{d x(t)}{dt}=v_\theta^x(z(t),t),
\qquad
\frac{d c(t)}{dt}=v_\theta^c(z(t),t).
\end{equation}

The resulting model supports the joint interpolation of molecular states and spatial tissue organization at unobserved time points.

\subsection{Structure-Aware Temporal Coupling}

To infer a probabilistic transport coupling between adjacent snapshots, we use fused Gromov--Wasserstein unbalanced optimal transport (FGWUOT). For snapshots
$\mathcal{S}_k$ and $\mathcal{S}_{k+1}$, we define the expression cost

\begin{equation}
C_{ij}^{x}=\left\| x_i^k-x_j^{k+1} \right\|_2^2,
\end{equation}

and the within-snapshot spatial distance matrices

\begin{equation}
D_{ii'}^k = \left\| c_i^k-c_{i'}^k \right\|_2,
\qquad
D_{jj'}^{k+1} = \left\| c_j^{k+1}-c_{j'}^{k+1} \right\|_2.
\end{equation}

Let $a^k\in\mathbb{R}_+^{n_k}$ and
$a^{k+1}\in\mathbb{R}_+^{n_{k+1}}$ denote the empirical cell masses. The transport plan is obtained by solving

\begin{align}
P_k^* = \arg\min_{P\geq 0}\;&(1-\alpha)\left\langle C^x,P\right\rangle\nonumber\\&+\alpha\sum_{i,i',j,j'}
\left(D_{ii'}^k-D_{jj'}^{k+1}\right)^2 P_{ij}P_{i'j'}\nonumber\\
&+\varepsilon H(P)+\rho_k\operatorname{KL}\left(
P\mathbf{1}\,\|\,a^k\right)\nonumber\\
&+\rho_{k+1}\operatorname{KL}\left(P^\top\mathbf{1}\,\|\,a^{k+1}\right),
\label{eq:fgwuot}
\end{align}

where $\alpha$ balances expression similarity and spatial structure,$\varepsilon$ controls entropic regularization, and $\rho_k$ and $\rho_{k+1}$ control deviations from the snapshot marginals. The unbalanced formulation allows the total transported mass to vary, thereby accommodating
cell proliferation, death, and sampling differences across time.

\subsection{Rigid Spatial Registration}

Although FGWUOT compares the internal geometry of two snapshots, their coordinate systems may differ because of global tissue rotation and translation. We therefore estimate a weighted rigid transformation from the
FGWUOT coupling:

\begin{equation}
\left(R_k^*,b_k^*\right)=
\arg\min_{\substack{R\in\mathrm{SO}(d_c)\\b\in\mathbb{R}^{d_c}}}
\sum_{i,j}P_{k,ij}^*\left\|
c_i^k-\left(R c_j^{k+1}+b\right)\right\|_2^2.
\label{eq:rigid_registration}
\end{equation}

Here, $R_k^*$ is the optimal $d_c$-dimensional rotation matrix and $b_k^*$ is the optimal translation vector. The coordinates of the later snapshot are then transformed into the coordinate system of the preceding snapshot:

\begin{equation}
\widetilde{c}_j^{k+1}
=
R_k^*c_j^{k+1}+b_k^*.
\end{equation}

For notational simplicity, we use $c_j^{k+1}$ to denote the registered coordinates in the following sections. The resulting transport plan provides probabilistic cross-time transition weights, while rigid registration removes technical
coordinate-system differences before learning spatial tissue dynamics.

\subsection{Coupling-Guided Joint Flow Matching}

We normalize the unbalanced transport plan to define a sampling distribution

\begin{equation}
\pi_k(i,j)=\frac{P_{k,ij}^*}{\sum_{i',j'}P_{k,i'j'}^*}.
\end{equation}

During training, we first sample an adjacent time interval
$[t_k,t_{k+1}]$ and then sample a source--target pair
$(z_0,z_1)$ according to $\pi_k$, where

\begin{equation}
z_0=(x_i^k,c_i^k),
\qquad
z_1=(x_j^{k+1},c_j^{k+1}).
\end{equation}

For $\tau\sim\mathcal{U}(0,1)$, we construct the linear interpolation

\begin{equation}
z_\tau=(1-\tau)z_0+\tau z_1,
\qquad
t_\tau=t_k+\tau(t_{k+1}-t_k).
\end{equation}

The corresponding conditional target velocities are

\begin{equation}
u_k^x=\frac{x_1-x_0}{t_{k+1}-t_k},
\qquad
u_k^c=\frac{c_1-c_0}{t_{k+1}-t_k}.
\end{equation}

The neural velocity field is trained to recover these coupling-induced
directions from interpolated cell states.

\subsubsection{Gene-Expression Velocity Loss}

We use a squared-error flow-matching objective for the gene-expression velocity:

\begin{equation}
\mathcal{L}_x(\theta)=\mathbb{E}_{\substack{
k,\,(z_0,z_1)\sim\pi_k\\\tau\sim\mathcal{U}(0,1)
}}\left[\left\|v_\theta^x(z_\tau,t_\tau)-\frac{x_1-x_0}{t_{k+1}-t_k}\right\|_2^2\right].
\label{eq:gene_fm}
\end{equation}

This objective encourages the predicted molecular velocity to follow the expression transitions implied by the FGWUOT coupling.

\subsubsection{Robust Spatial Velocity Loss}

Coordinate transition targets derived from sampled cross-time pairs are typically noisier than gene-expression transition targets because of tissue deformation, imperfect sectioning, and registration uncertainty. We therefore use the smooth $\ell_1$ loss instead of a squared-error loss for spatial velocity learning:

\begin{equation}
\mathcal{L}_c(\theta)
=
\mathbb{E}_{\substack{
k,\;(\mathbf{z}_0,\mathbf{z}_1)\sim\pi_k\\
\tau\sim\mathcal{U}(0,1)
}}
\left[
\operatorname{SmoothL1}_{\beta}
\left(
v_\theta^c(\mathbf{z}_\tau,t_\tau)
-
\mathbf{u}_k^c
\right)
\right],
\label{eq:coordinate_velocity_loss}
\end{equation}

where

\begin{equation}
\operatorname{SmoothL1}_{\beta}(r)
=
\begin{cases}
\dfrac{r^2}{2\beta},
& |r| < \beta, \\[6pt]
|r|-\dfrac{\beta}{2},
& |r| \geq \beta.
\end{cases}
\label{eq:smooth_l1}
\end{equation}

The Smooth $\ell_1$ penalty is applied element-wise to the coordinate velocity residuals and averaged over all spatial dimensions. We set $\beta=0.1$ in all experiments.

\subsubsection{Coordinate Distribution Matching}

Pairwise velocity supervision alone does not guarantee that the generated coordinates recover the global geometry of the target tissue. We therefore introduce a coordinate distribution matching (CDM) loss.

Starting from the source coordinates at $t_k$, we integrate the learned coordinate velocity field to obtain the predicted target coordinate set

\begin{equation}
\widehat{\mathcal{C}}_{k+1}
=
\left\{
\widehat{c}_i(t_{k+1})
\right\}_{i=1}^{\widehat n_{k+1}}.
\end{equation}

Let

\begin{equation}
\mathcal{C}_{k+1}
=
\left\{
c_j^{k+1}
\right\}_{j=1}^{n_{k+1}}
\end{equation}

denote the observed target coordinates after rigid registration. We measure
their discrepancy using the symmetric Chamfer distance:

\begin{align}
\mathcal{L}_{\mathrm{CDM}}(\theta)
=&
\frac{1}{|\widehat{\mathcal{C}}_{k+1}|}
\sum_{\widehat c\in\widehat{\mathcal{C}}_{k+1}}
\min_{c\in\mathcal{C}_{k+1}}
\left\|
\widehat c-c
\right\|_2
\nonumber\\
&+
\frac{1}{|\mathcal{C}_{k+1}|}
\sum_{c\in\mathcal{C}_{k+1}}
\min_{\widehat c\in\widehat{\mathcal{C}}_{k+1}}
\left\|
c-\widehat c
\right\|_2.
\label{eq:cdm}
\end{align}

The first term penalizes generated coordinates outside the target tissue,
whereas the second term encourages the generated coordinates to cover the
complete target structure.

\subsubsection{Overall Training Objective}

The complete reconstruction objective is

\begin{equation}
\mathcal{L}_{\mathrm{FM}}
=
\lambda_x\mathcal{L}_x
+
\lambda_c\mathcal{L}_c
+
\lambda_{\mathrm{CDM}}\mathcal{L}_{\mathrm{CDM}},
\label{eq:total_loss}
\end{equation}

where $\lambda_x$, $\lambda_c$, and $\lambda_{\mathrm{CDM}}$ balance
gene-expression reconstruction, local spatial velocity learning, and global
coordinate distribution matching, respectively.

\subsection{Parallel Differentiation-Potential Learning}

The gene-spatial velocity field $v_\theta$
is designed to flexibly model gene-expression and spatial dynamics for continuous interpolation. However, its local, high-dimensional velocities do not directly assign a single, comparable value to each cell state along the inferred differentiation progression. We therefore independently learn a time-conditioned potential function

\begin{equation}
U_\phi:
\mathbb{R}^{d_x+d_c}\times\mathbb{R}
\rightarrow
\mathbb{R}.
\end{equation}

Its negative gradient is trained to follow the same earlier-to-later transition directions, thereby defining the auxiliary gradient field

\begin{equation}
v_\phi^{\mathrm{pot}}(z,t)
=
-\nabla_z U_\phi(z,t)
=
\left(
-\nabla_xU_\phi(z,t),
-\nabla_cU_\phi(z,t)
\right).
\end{equation}

The reconstruction and potential branches are parameterized by separate neural networks and optimized independently. They share the FGWUOT-derived source–target pairs and conditional velocity supervision, but no equality is imposed between $v_\theta$ and $v_\phi^{pot}$.

The potential branch uses analogous coupling-guided velocity-matching objectives for its gene-expression and spatial gradients. Its training objective is

\begin{equation}
    \mathcal{L}_{pot}=\lambda_x\mathcal{L}_x^{pot}+\lambda_c\mathcal{L}_c^{pot},
\end{equation}

where

\begin{equation}
    \mathcal{L}_x^{pot}=\mathbb{E}[||-\nabla_xU_\phi(z_\tau,t_\tau)-u^k_x||^2_2],
\end{equation}

and

\begin{equation}
    \mathcal{L}_c^{pot}=\mathbb{E}[SmoothL1_\beta(-\nabla_cU_\phi(z_\tau,t_\tau)-u^k_c)].
\end{equation}

Unlike the reconstruction objective, the potential objective does not include coordinate distribution matching, because the potential branch is used for potential landscape interpretation rather than spatial reconstruction.

Because $-\nabla_zU_\phi$ is trained to follow transition velocities from earlier to later snapshots, the learned potential is encouraged to decrease locally along data-supported transitions. We therefore use its values to compare and organize cell states along inferred differentiation progressions, primarily among cells evaluated at the same developmental time.

\subsection{Continuous-Time Inference}

Given cells observed at time $t_k$, we predict their states at an arbitrary
time $t\in[t_k,t_{k+1}]$ by solving

\begin{equation}
\widehat z(t)
=
z(t_k)
+
\int_{t_k}^{t}
v_\theta\left(\widehat z(s),s\right)\,ds.
\end{equation}

Equivalently, the molecular and spatial predictions are

\begin{align}
\widehat x(t)
&=
x(t_k)
+
\int_{t_k}^{t}
v_\theta^x\left(\widehat z(s),s\right)\,ds,\\
\widehat c(t)
&=
c(t_k)
+
\int_{t_k}^{t}
v_\theta^c\left(\widehat z(s),s\right)\,ds.
\end{align}

This joint integration produces gene-expression profiles and spatial
coordinates at unobserved intermediate time points. For predictions spanning
multiple observed intervals, the output at $t_{k+1}$ is used as the initial
state for the subsequent interval.

\begin{table*}[t]
\centering
\caption{
Interpolation performance across time-resolved spatial transcriptomic
datasets. Lower values indicate better performance. The best and second-best
results are highlighted in bold and underlined, respectively. Spatial metrics
are reported only for methods that generate spatial coordinates.
}
\label{tab:interpolation_results}

{\scriptsize
\setlength{\tabcolsep}{3.5pt}
\renewcommand{\arraystretch}{0.95}

\begin{tabular*}{0.92\textwidth}{
@{\extracolsep{\fill}}
ll ccc c cccccc
}
\toprule
& &
\multicolumn{3}{c}{ARTISTA} &
\multicolumn{1}{c}{MOSTA} &
\multicolumn{6}{c}{PRISTA4D} \\
\cmidrule(lr){3-5}
\cmidrule(lr){6-6}
\cmidrule(lr){7-12}

Metric
& Method
& 5DPI
& 10DPI
& 15DPI
& E10.5
& 12hpa
& 36hpa
& 3dpa
& 5dpa
& 7dpa
& 10dpa \\

\midrule
\multicolumn{12}{@{}l}{\textit{Gene Expression}} \\
\multirow[t]{4}{*}{MMD $\downarrow$}
& ST-FateFlow
& \textbf{0.0147}
& \textbf{0.0344}
& \textbf{0.0557}
& \textbf{0.0079}
& \textbf{0.0295}
& \textbf{0.0290}
& \textbf{0.0277}
& \textbf{0.0140}
& \textbf{0.0190}
& \textbf{0.0296} \\

& stVCR
& 0.1011
& 0.1246
& 0.1540
& 0.1271
& 0.0915
& 0.0881
& 0.0864
& \underline{0.0457}
& \underline{0.0411}
& \underline{0.0397} \\

& PRESCIENT
& \underline{0.0830}
& \underline{0.1013}
& \underline{0.1251}
& \underline{0.0936}
& \underline{0.0510}
& \underline{0.0523}
& 0.0613
& 0.0689
& 0.0669
& 0.0517 \\

& scNODE
& 0.1110
& 0.1203
& 0.1585
& 0.1392
& 0.0739
& 0.0553
& \underline{0.0555}
& 0.0896
& 0.1593
& 0.3282 \\

\cmidrule(lr){2-12}

\multirow[t]{4}{*}{$W_1$ $\downarrow$}
& ST-FateFlow
& \underline{0.8773}
& \underline{0.8604}
& \underline{0.8444}
& \textbf{0.6769}
& \underline{0.6944}
& \underline{0.6922}
& \underline{0.7093}
& \underline{0.7599}
& \underline{0.7828}
& 0.7782 \\

& stVCR
& \textbf{0.8650}
& \textbf{0.8037}
& \textbf{0.8206}
& \underline{0.7221}
& \textbf{0.5939}
& \textbf{0.6264}
& \textbf{0.6140}
& \textbf{0.7148}
& \textbf{0.7203}
& \textbf{0.7224} \\

& PRESCIENT
& 0.8783
& 0.8913
& 0.9074
& 0.9275
& 0.7728
& 0.7733
& 0.7835
& 0.7948
& 0.8091
& \underline{0.7705} \\

& scNODE
& 0.9076
& 0.9017
& 0.9306
& 0.9397
& 0.8032
& 0.7769
& 0.7836
& 0.8186
& 0.8724
& 0.9132 \\

\midrule
\multicolumn{12}{@{}l}{\textit{Spatial Coordinate}} \\
\multirow[t]{4}{*}{Chamfer $\downarrow$}
& ST-FateFlow
& \underline{0.0226}
& \textbf{0.0228}
& \textbf{0.0212}
& \textbf{0.0134}
& \textbf{0.0900}
& \textbf{0.0564}
& \textbf{0.0475}
& \textbf{0.0295}
& \textbf{0.0216}
& \textbf{0.0257} \\

& stVCR
& \textbf{0.0200}
& \underline{0.0246}
& \underline{0.0239}
& \underline{0.0962}
& \underline{1.0499}
& \underline{0.8690}
& \underline{1.0753}
& \underline{1.0139}
& \underline{1.1224}
& \underline{1.2256} \\

\cmidrule(lr){2-12}

\multirow[t]{4}{*}{$W_1$ $\downarrow$}
& ST-FateFlow
& \underline{0.0889}
& \textbf{0.0674}
& \underline{0.0883}
& \textbf{0.0377}
& \textbf{0.0748}
& \textbf{0.0682}
& \textbf{0.0584}
& \textbf{0.0518}
& \textbf{0.0440}
& \textbf{0.0432} \\

& stVCR
& \textbf{0.0669}
& \underline{0.0692}
& \textbf{0.0831}
& \underline{0.0774}
& \underline{0.2363}
& \underline{0.1955}
& \underline{0.2667}
& \underline{0.3190}
& \underline{0.3246}
& \underline{0.4090} \\

\cmidrule(lr){2-12}

\multirow[t]{4}{*}{HD95 $\downarrow$}
& ST-FateFlow
& \underline{0.0614}
& \textbf{0.0524}
& \textbf{0.0509}
& \textbf{0.0425}
& \textbf{0.2287}
& \textbf{0.1125}
& \textbf{0.1135}
& \textbf{0.0576}
& \textbf{0.0313}
& \textbf{0.0481} \\

& stVCR
& \textbf{0.0413}
& \underline{0.0662}
& \underline{0.0658}
& \underline{0.2493}
& \underline{0.9309}
& \underline{1.2876}
& \underline{1.0692}
& \underline{1.1063}
& \underline{1.1381}
& \underline{1.0647} \\

\bottomrule
\end{tabular*}
}
\end{table*}

\section{Experiments}
\subsection{Experimental Setup}

\paragraph{Datasets.}
We evaluate ST-FateFlow on three single-cell resolution, time-resolved spatial transcriptomic datasets: the Mouse Organogenesis Spatiotemporal Transcriptomic Atlas (MOSTA) \cite{chen2022spatiotemporal}, the Planarian
Regenerative Interactive Spatiotemporal Transcriptomic Atlas in four dimensions (PRISTA4D) \cite{han20264d}, and the Axolotl Regenerative
Telencephalon Interpretation via Spatiotemporal Transcriptomic Atlas (ARTISTA) \cite{wei2022single}. These datasets cover mouse embryonic organogenesis, planarian regeneration, and axolotl brain development and
regeneration, providing diverse settings for evaluating continuous gene-expression and spatial dynamics. 

\paragraph{Preprocessing.}
For each dataset, we remove low-quality cells, apply library-size normalization followed by a $\log(1+x)$ transformation, and select the 2,000 most highly variable genes. The same selected genes are retained across all
time points within each dataset to ensure a consistent expression space. ST-FateFlow directly models these gene-expression profiles together with their two- or three-dimensional spatial coordinates and does not require PCA or other dimensionality-reduction preprocessing.

\paragraph{Baselines.}
We compare ST-FateFlow with stVCR , PRESCIENT, and scNODE. Since few existing methods reconstruct continuous gene-expression and spatial dynamics jointly, stVCR serves as the primary spatiotemporal baseline. PRESCIENT and scNODE provide representative comparisons with continuous
time-series scRNA-seq models. All methods are evaluated using the same temporal splits and selected genes. Gene-expression metrics are reported for all methods, while spatial-coordinate metrics are reported only for ST-FateFlow and stVCR because PRESCIENT and scNODE do not generate spatial coordinates.

\paragraph{Implementation Details.}
We evaluate interpolation performance by comparing predicted cells with the observed cells at each held-out time point. For gene-expression reconstruction, we report the maximum mean discrepancy with an RBF kernel (MMD) and the 1-Wasserstein distance ($W_1$). MMD evaluates the discrepancy between predicted and observed expression distributions in kernel space, while $W_1$ measures
the minimum transport cost required to match the two distributions. For spatial-coordinate reconstruction, we report the symmetric Chamfer distance, the spatial 1-Wasserstein distance, and the 95th-percentile Hausdorff distance (HD95). Chamfer distance measures the average bidirectional nearest-neighbor discrepancy between predicted and observed coordinates. HD95 captures
large spatial deviations while remaining less sensitive to isolated outliers than the standard Hausdorff distance. Lower values indicate better performance for all evaluation metrics.

\subsection{Quantitative Comparison}

We first evaluate whether the learned joint velocity field can accurately reconstruct gene-expression distributions and spatial tissue organization at unobserved time points. We hold out selected intermediate snapshots from training and compare the interpolated cells against the corresponding observations. The quantitative results across ten interpolation settings from ARTISTA, MOSTA, and PRISTA4D are summarized in Table~\ref{tab:interpolation_results}.

Because cells observed in different spatial transcriptomic snapshots are unpaired, expression interpolation is more appropriately evaluated at the population level than using cell-wise errors. We therefore use MMD as the primary expression metric because it measures the discrepancy between predicted and observed expression distributions without requiring explicit cell matching. ST-FateFlow achieves the lowest MMD in all ten interpolation settings. For ARTISTA, it reduces MMD by $82.3\%$, $66.0\%$, and $55.5\%$ at 5, 10, and 15 DPI, respectively, relative to the corresponding second-best method. For MOSTA at E10.5, it reduces MMD from
$0.0936$ to $0.0079$, corresponding to a $91.6\%$ improvement. ST-FateFlow also consistently achieves the lowest MMD across all six PRISTA4D time points,
with reductions ranging from $25.4\%$ to $69.3\%$ over the second-best results. Although stVCR generally obtains lower expression-space Wasserstein distances, ST-FateFlow achieves the best result on MOSTA and the second-best result in eight of the remaining nine settings. Overall, the consistently lower MMD indicates that the learned gene-expression velocity field more accurately transports source populations toward the observed intermediate expression distributions, suggesting that the inferred gene-expression dynamics better capture the population-level transcriptional changes associated with development and regeneration.

ST-FateFlow also demonstrates strong performance in reconstructing spatial tissue organization, achieving the best result in 26 of the 30 dataset--time--metric combinations. PRISTA4D provides a particularly challenging setting because it requires the reconstruction of continuous three-dimensional spatial dynamics rather than two-dimensional tissue sections. Nevertheless, ST-FateFlow outperforms stVCR on all three spatial metrics across all six PRISTA4D time points. Specifically, it reduces Chamfer
distance by $91.4\%$--$98.1\%$, spatial Wasserstein distance by $65.1\%$--$89.4\%$, and HD95 by $75.4\%$--$97.2\%$. ST-FateFlow also outperforms stVCR on all spatial metrics for ARTISTA at 10 DPI and MOSTA at
E10.5, while obtaining lower Chamfer distance and HD95 for ARTISTA at 15 DPI. Although stVCR performs better for ARTISTA at 5 DPI and achieves a slightly lower spatial Wasserstein distance at 15 DPI, ST-FateFlow is substantially more consistent across datasets and interpolation times. These results suggest that combining robust spatial-velocity supervision with coordinate distribution matching improves both local coordinate accuracy and global tissue-structure preservation in two- and three-dimensional tissues.

\begin{figure}[t]
    \centering

    \makebox[\columnwidth][l]{\scriptsize\textbf{(a)}}
    \par\vspace{0.5mm}%
    \makebox[\columnwidth][c]{%
        \begin{minipage}[b]{0.414\columnwidth}
            \centering
            \raisebox{1.5mm}{%
                \includegraphics[width=\linewidth]
                    {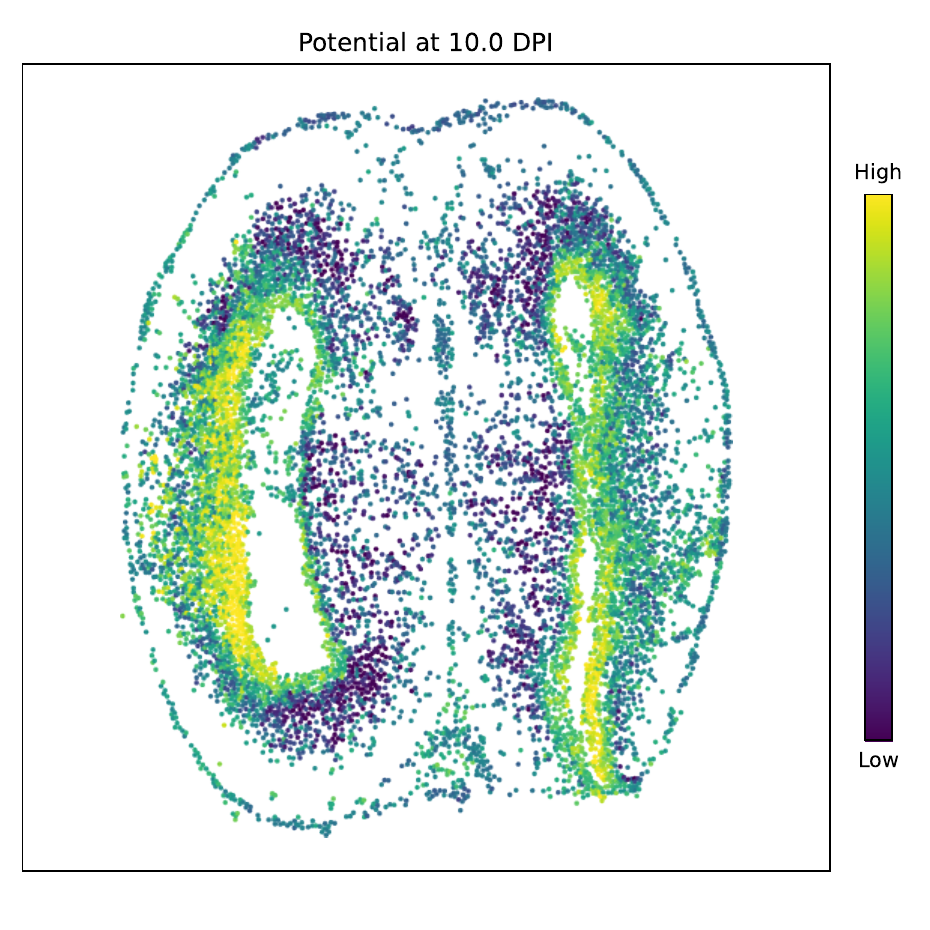}%
            }
        \end{minipage}%
        \hspace{0.085\columnwidth}%
        \begin{minipage}[b]{0.468\columnwidth}
            \centering
            \includegraphics[width=\linewidth]
                {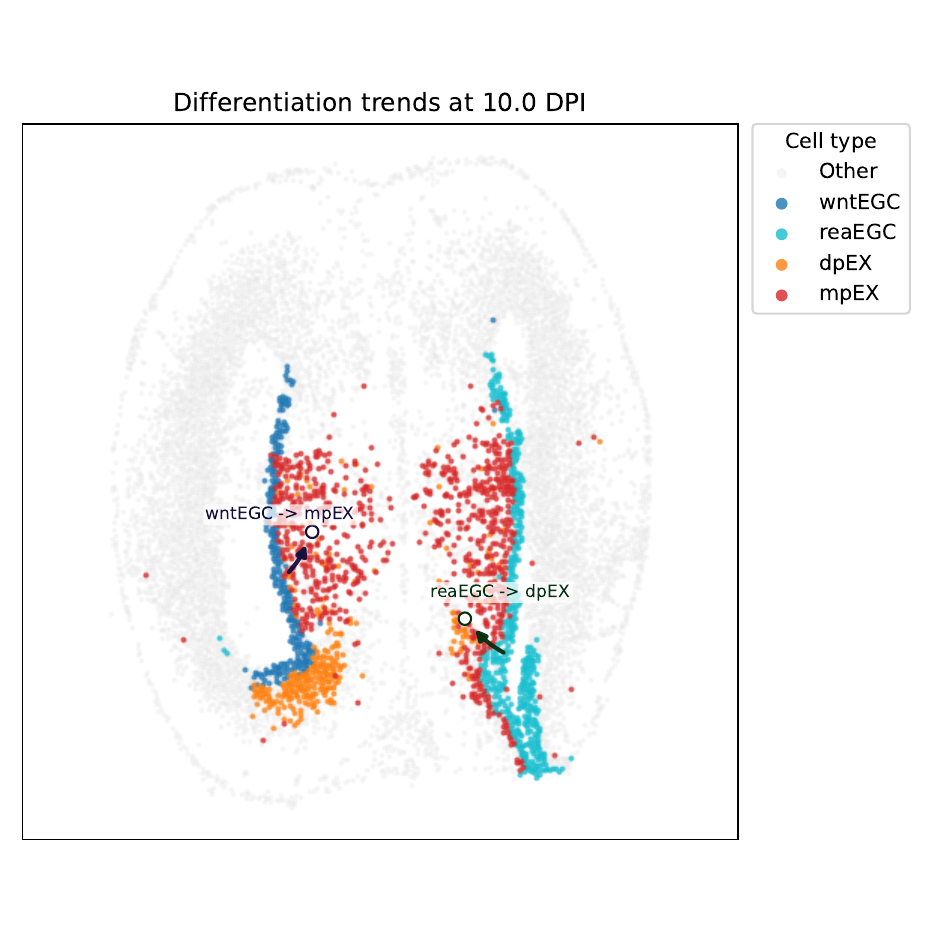}
        \end{minipage}%
    }
    \makebox[\columnwidth][l]{\scriptsize\textbf{(b)}}
    \par\vspace{0.5mm}
    \includegraphics[width=0.96\columnwidth]
        {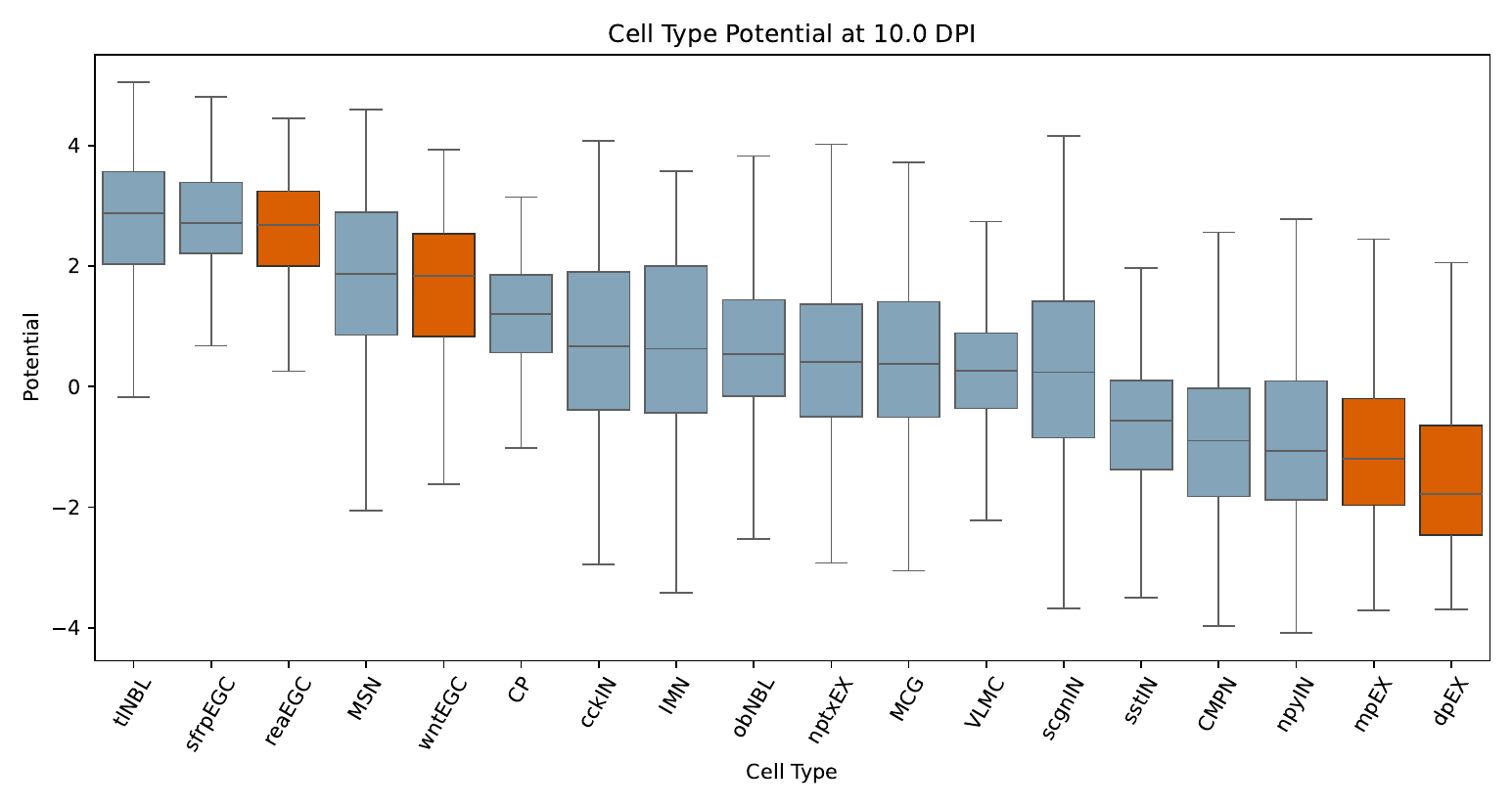}

    \caption{
    Analysis of the learned potential landscape at 10 DPI.
    (a) Spatial distribution of the learned potential and reported differentiation trajectories.
    (b) Cell-type potential distributions at 10.0 DPI; red boxes denote the cell types of interest that exhibit clear differentiation trajectories.
    }
    \label{fig:potential_analysis}
\end{figure}


\begin{figure*}[t]
    \centering
    \begin{subfigure}[t]{0.42\linewidth}
        \vspace{0pt}%
        \includegraphics[width=\linewidth]{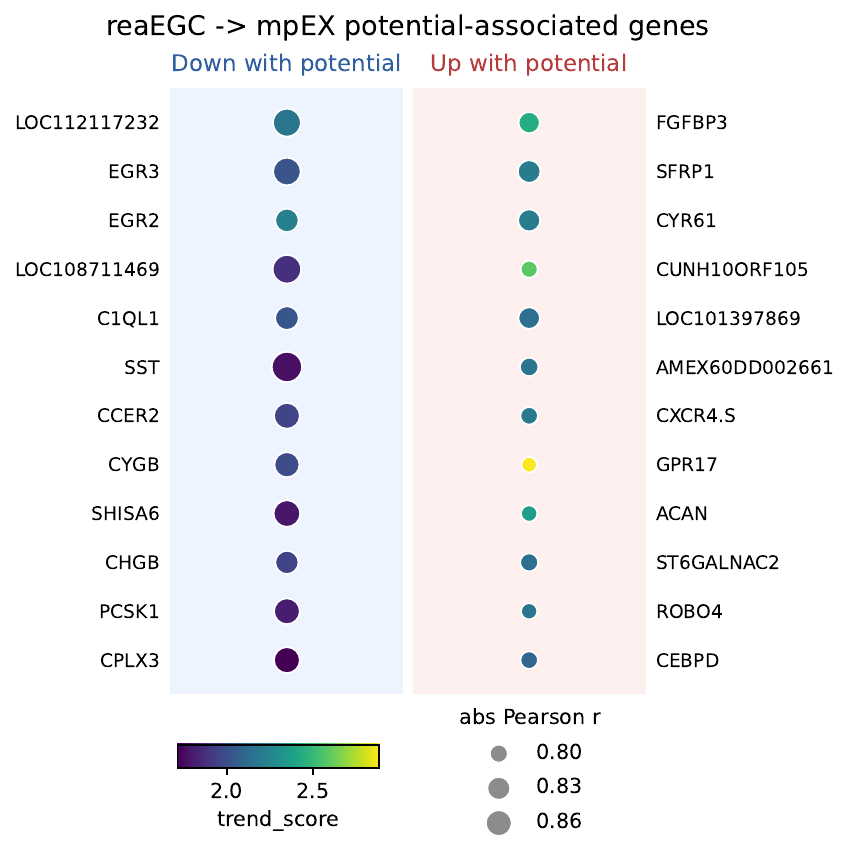}
    \end{subfigure}%
    \hfill
    \begin{subfigure}[t]{0.48\linewidth}
        \vspace{0pt}%
        \includegraphics[width=\linewidth]{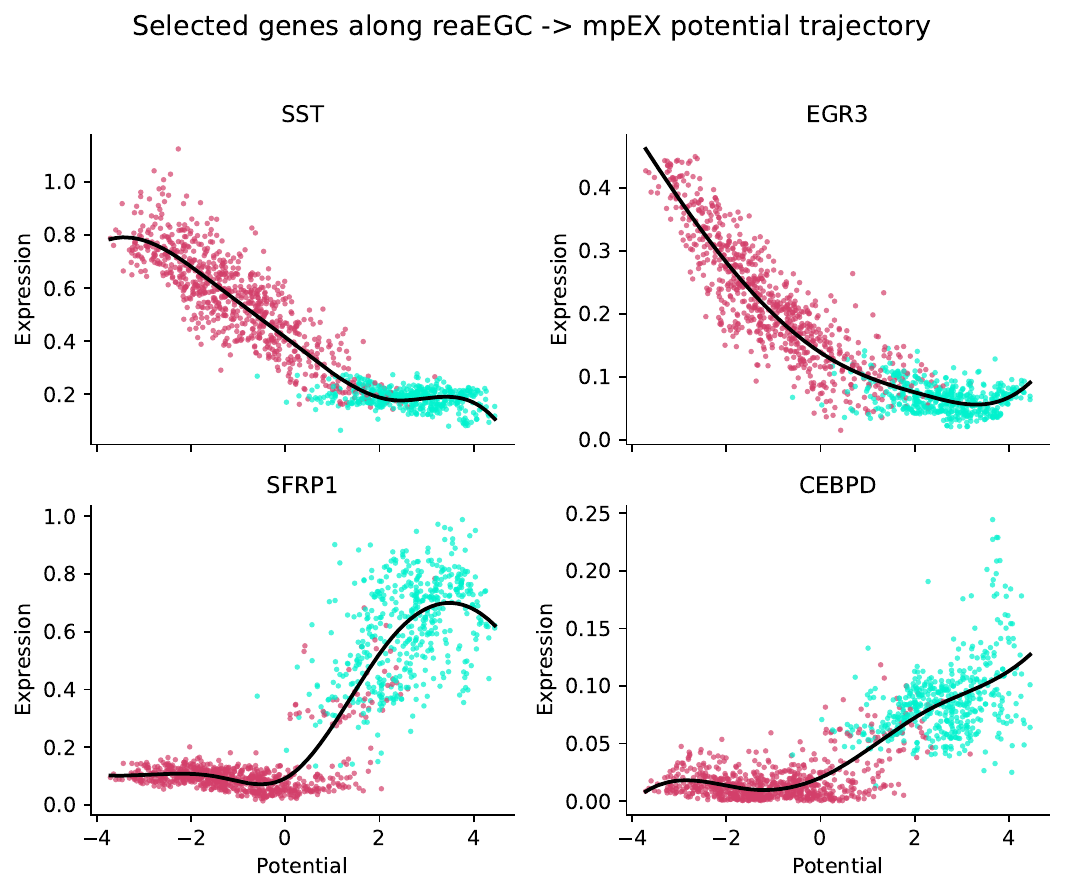}
    \end{subfigure}
    \caption{
        Potential-associated gene-expression patterns across reaEGC and mpEX cells at 10 DPI. 
        (left) Genes exhibiting strong decreasing or increasing trends with the learned potential. Dot size represents the absolute Pearson correlation between gene expression and potential, while color represents the trend score combining correlation strength with the expression fold change between high- and low-potential cells. 
        (right) Expression trends of selected genes along the potential landscape. Points represent individual mpEX (pink) and reaEGC (cyan) cells, and black curves show smoothed expression trends. \textit{SST} and \textit{EGR3} decrease with potential, whereas \textit{SFRP1} and \textit{CEBPD} increase with potential.
    }
    \label{fig:potential_genes}
\end{figure*}

\begin{figure*}[t]
    \centering
    \includegraphics[width=0.96\textwidth]
    {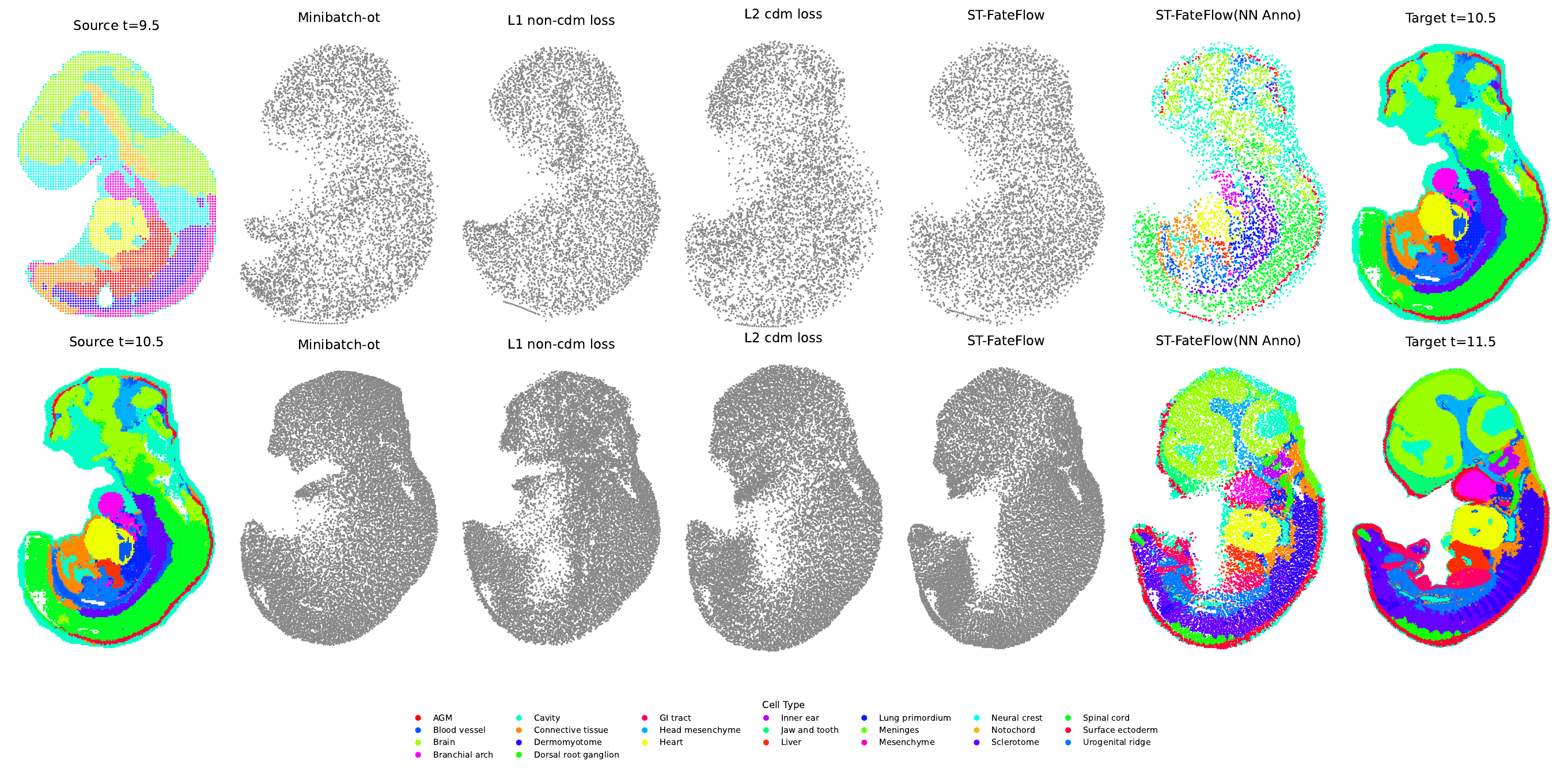}
    \caption{Qualitative ablation study of spatial interpolation on the MOSTA dataset. The first and last columns show the observed source and target slices, respectively. The intermediate columns show predictions obtained by replacing FGWUOT with minibatch optimal transport, using the $\ell_1$ coordinate loss without CDM, using the mean squared error loss with CDM, and using the complete ST-FateFlow objective with Smooth $\ell_1$ and CDM. The ST-FateFlow (NN Anno) column displays the complete-model prediction colored using cell-type labels transferred from the nearest cells in the target slice. }
    \label{fig:qualitative_comparison}
\end{figure*}

\subsection{Potential Landscape Analysis}

We analyze whether the learned potential landscape is consistent with previously reported differentiation progressions. Figure~\ref{fig:potential_analysis}(a)
shows the spatial distribution of the learned potential at 10 DPI together with cell populations involved in the reported wntEGC $\rightarrow$ mpEX and reaEGC $\rightarrow$ rIPC2 $\rightarrow$ dpEX paths. As shown in
Figure~\ref{fig:potential_analysis}(b), the progenitor-like wntEGC and reaEGC populations exhibit relatively high potential, whereas the differentiated mpEX and dpEX populations occupy lower-potential states. For the longer reaEGC $\rightarrow$ rIPC2 $\rightarrow$ dpEX progression, rIPC2 exhibits an intermediate potential between reaEGC and dpEX, as shown in Appendix Figure~\ref{fig:appendix_potential_boxplots}. Together, these results agree with the expected progression from EGC populations toward differentiated neuronal states, suggesting that the learned potential landscape captures biologically meaningful differences along differentiation progression.

We next characterize transcriptional changes associated with the learned potential at 10 DPI. Rather than preselecting marker genes, we screen expressed genes across reaEGC and mpEX cells using a trend score that jointly considers the Pearson correlation between gene expression and potential and the fold change between high- and low-potential cells. As shown in Figure~\ref{fig:potential_genes}, genes with strong potential-associated trends are separated according to whether their expression increases or decreases with potential. Point size represents the absolute Pearson correlation, whereas color indicates the trend score. Genes decreasing with potential include several genes related to neuronal or synaptic functions, such as \textit{SST}, \textit{C1QL1}, \textit{CHGB}, \textit{PCSK1}, and \textit{CPLX3}. Because differentiation is associated with progression from high to low potential, these genes become more highly expressed toward the low-potential neuronal state, consistent with the acquisition of neuronal transcriptional features. Conversely, genes such as \textit{SFRP1}, \textit{CEBPD}, \textit{CXCR4}, and \textit{GPR17} increase with potential and are preferentially expressed in the high-potential reaEGC population.

Among these genes, we further examine \textit{SFRP1} and \textit{CEBPD}, which were identified through systematic screening but were not emphasized in previous analyses of this dataset or previously implicated in this specific axolotl cell-state transition. \textit{SFRP1} is a modulator of Wnt signaling and has been associated with the regulation of neural progenitor maintenance, activation, and differentiation \cite{donega2022single}. \textit{CEBPD} is an injury- and inflammation-responsive transcription factor associated with reactive glial responses following neural injury \cite{wang2016astrocytic}. Their preferential expression in high-potential reaEGC cells supports their consideration as biologically plausible candidates associated with progenitor or regenerative states. These associations do not establish direct regulatory or causal roles in axolotl neuronal differentiation, but illustrate how the learned potential can support the discovery of candidate genes associated with cell-state progression.

\subsection{Ablation Study}

We conduct qualitative ablation experiments on MOSTA to examine the contributions of the cross-time transport coupling and the spatial training objective. First, we replace FGWUOT with minibatch optimal transport while retaining the remaining training procedure. Minibatch OT provides a memory-efficient alternative but estimates transport couplings independently within local cell subsets, which can reduce the consistency of the reconstructed global tissue structure. As shown in Figure~\ref{fig:qualitative_comparison}, this variant captures the coarse target geometry but introduces locally inconsistent densities and structural discontinuities, particularly during the 10.5-to-11.5 interpolation.

We further compare three spatial training objectives. Using the $\ell_1$ coordinate loss without CDM preserves the approximate tissue outline but produces uneven local densities and visible gaps, indicating that point-wise regression alone does not sufficiently constrain the predicted population-level geometry. Combining the mean squared error (MSE) loss with CDM improves global distribution matching, but the resulting coordinates remain locally contracted or overly smoothed in regions undergoing substantial deformation. In contrast, the complete ST-FateFlow objective, which combines Smooth $\ell_1$ regression with CDM, provides a better balance between point-wise coordinate accuracy and population-level structure preservation. The nearest-neighbor annotations further show that the predicted coordinates retain coherent spatial organization across major tissue regions. These annotations are transferred from the target slice solely for visualization and are not used for model training or coordinate prediction. Overall, the comparison supports the complementary contributions of structure-aware FGWUOT coupling, robust coordinate regression, and distribution-level spatial matching.

\section{Conclusion}

We introduced ST-FateFlow, a unified framework for reconstructing continuous spatiotemporal dynamics and learning time-conditioned differentiation-associated potential landscapes from sparsely sampled transcriptomic snapshots. ST-FateFlow uses FGWUOT to establish structure-aware transport coupling across unpaired observations and jointly learns gene-expression and spatial-coordinate velocity fields through flow matching. Experiments on multiple 2D and 3D spatiotemporal datasets demonstrate that the learned dynamics accurately capture changes in both molecular states and tissue organization, with consistent improvements in distributional and spatial interpolation metrics. In addition, the time-conditioned potential recapitulates the expected high-to-low ordering along previously established differentiation paths, including wntEGC $\rightarrow$ mpEX and reaEGC $\rightarrow$ rIPC2 $\rightarrow$ dpEX, and supports the identification of potential-associated genes such as \textit{CEBPD} and \textit{SFRP1}. Together, ST-FateFlow provides an integrated approach for predicting continuous gene-spatial dynamics and characterizing biologically meaningful cell-state organization in time-resolved transcriptomic data.

\section{Limitations and Ethical Considerations}

ST-FateFlow depends on the quality of cross-time coupling and spatial registration and may be affected by batch effects or missing intermediate states. Learning the potential requires additional gradient computation over high-dimensional gene-spatial states, increasing computational cost. The learned potential should be interpreted as a relative organization of cell states along inferred differentiation progressions, rather than as a fate probability or physical energy. This study uses only publicly available datasets and involves no new human-subject or animal experiments.

\section{Generative AI Usage}
In this work, we mainly employ ChatGPT to identify and correct syntax errors and polish writing.

\begin{acks}
This research is supported by the National Natural Science Foundation of China (Grant No. W2431045) and Fundamental and Interdisciplinary Disciplines Breakthrough Plan of the Ministry of Education of China (No. JYB2025XDXM118).
\end{acks}

\bibliographystyle{ACM-Reference-Format}
\bibliography{sample-base}

@String{Computing = "Computing" }

@String{Springer = "Springer-Verlag" }

@article{gerri2020human,
  title={Human embryogenesis: a comparative perspective},
  author={Gerri, Claudia and Menchero, Sergio and Mahadevaiah, Shantha K and Turner, James MA and Niakan, Kathy K},
  journal={Annual review of cell and developmental biology},
  volume={36},
  number={1},
  pages={411--440},
  year={2020},
  publisher={Annual Reviews}
}

@article{liu2024spatiotemporal,
  title={Spatiotemporal omics for biology and medicine},
  author={Liu, Longqi and Chen, Ao and Li, Yuxiang and Mulder, Jan and Heyn, Holger and Xu, Xun},
  journal={Cell},
  volume={187},
  number={17},
  pages={4488--4519},
  year={2024},
  publisher={Elsevier}
}

@article{lazar2026spatial,
  title={Spatial architecture of development and disease},
  author={L{\'a}z{\'a}r, Enik{\H{o}} and Lundeberg, Joakim},
  journal={Nature Reviews Genetics},
  volume={27},
  number={2},
  pages={118--136},
  year={2026},
  publisher={Nature Publishing Group UK London}
}

@article{lipman2022flow,   
title={Flow matching for generative modeling},   author={Lipman, Yaron and Chen, Ricky TQ and Ben-Hamu, Heli and Nickel, Maximilian and Le, Matt},   journal={arXiv preprint arXiv:2210.02747},   year={2022} 
}

@article{la2018rna,
  title={RNA velocity of single cells},
  author={La Manno, Gioele and Soldatov, Ruslan and Zeisel, Amit and Braun, Emelie and Hochgerner, Hannah and Petukhov, Viktor and Lidschreiber, Katja and Kastriti, Maria E and L{\"o}nnerberg, Peter and Furlan, Alessandro and others},
  journal={Nature},
  volume={560},
  number={7719},
  pages={494--498},
  year={2018},
  publisher={Nature Publishing Group UK London}
}

@article{bergen2020generalizing,
  title={Generalizing RNA velocity to transient cell states through dynamical modeling},
  author={Bergen, Volker and Lange, Marius and Peidli, Stefan and Wolf, F Alexander and Theis, Fabian J},
  journal={Nature biotechnology},
  volume={38},
  number={12},
  pages={1408--1414},
  year={2020},
  publisher={Nature Publishing Group US New York}
}

@article{wang2016astrocytic,
  title={Astrocytic CCAAT/enhancer-binding protein delta contributes to glial scar formation and impairs functional recovery after spinal cord injury},
  author={Wang, Shao-Ming and Hsu, Jung-Yu C and Ko, Chiung-Yuan and Chiu, Nai-En and Kan, Wai-Ming and Lai, Ming-Derg and Wang, Ju-Ming},
  journal={Molecular Neurobiology},
  volume={53},
  number={9},
  pages={5912--5927},
  year={2016},
  publisher={Springer}
}

@article{donega2022single,
  title={Single-cell profiling of human subventricular zone progenitors identifies SFRP1 as a target to re-activate progenitors},
  author={Donega, Vanessa and van der Geest, Astrid T and Sluijs, Jacqueline A and van Dijk, Roland E and Wang, Chi Chiu and Basak, Onur and Pasterkamp, R Jeroen and Hol, Elly M},
  journal={Nature communications},
  volume={13},
  number={1},
  pages={1036},
  year={2022},
  publisher={Nature Publishing Group UK London}
}

@article{tong2023conditional,
  title={Conditional flow matching: Simulation-free dynamic optimal transport},
  author={Tong, Alexander and Malkin, Nikolay and Huguet, Guillaume and Zhang, Yanlei and Rector-Brooks, Jarrid and Fatras, Kilian and Wolf, Guy and Bengio, Yoshua},
  journal={arXiv preprint arXiv:2302.00482},
  volume={2},
  number={3},
  year={2023}
}

@article{schiebinger2019optimal,
  title={Optimal-transport analysis of single-cell gene expression identifies developmental trajectories in reprogramming},
  author={Schiebinger, Geoffrey and Shu, Jian and Tabaka, Marcin and Cleary, Brian and Subramanian, Vidya and Solomon, Aryeh and Gould, Joshua and Liu, Siyan and Lin, Stacie and Berube, Peter and others},
  journal={Cell},
  volume={176},
  number={4},
  pages={928--943},
  year={2019},
  publisher={Elsevier}
}

@article{yeo2021generative,
  title={Generative modeling of single-cell time series with PRESCIENT enables prediction of cell trajectories with interventions},
  author={Yeo, Grace Hui Ting and Saksena, Sachit D and Gifford, David K},
  journal={Nature communications},
  volume={12},
  number={1},
  pages={3222},
  year={2021},
  publisher={Nature Publishing Group UK London}
}

@article{zhang2024scnode,
  title={scNODE: generative model for temporal single cell transcriptomic data prediction},
  author={Zhang, Jiaqi and Larschan, Erica and Bigness, Jeremy and Singh, Ritambhara},
  journal={Bioinformatics},
  volume={40},
  number={Supplement\_2},
  pages={ii146--ii154},
  year={2024},
  publisher={Oxford University Press}
}

@article{lange2022cellrank,
  title={CellRank for directed single-cell fate mapping},
  author={Lange, Marius and Bergen, Volker and Klein, Michal and Setty, Manu and Reuter, Bernhard and Bakhti, Mostafa and Lickert, Heiko and Ansari, Meshal and Schniering, Janine and Schiller, Herbert B and others},
  journal={Nature methods},
  volume={19},
  number={2},
  pages={159--170},
  year={2022},
  publisher={Nature Publishing Group US New York}
}

@article{shen2025inferring,
  title={Inferring cell trajectories of spatial transcriptomics via optimal transport analysis},
  author={Shen, Xunan and Zuo, Lulu and Ye, Zhongfei and Yuan, Zhongyang and Huang, Ke and Li, Zeyu and Yu, Qichao and Zou, Xuanxuan and Wei, Xiaoyu and Xu, Ping and others},
  journal={Cell systems},
  volume={16},
  number={2},
  year={2025},
  publisher={Elsevier}
}

@article{qiu2024spatiotemporal,
  title={Spatiotemporal modeling of molecular holograms},
  author={Qiu, Xiaojie and Zhu, Daniel Y and Lu, Yifan and Yao, Jiajun and Jing, Zehua and Min, Kyung Hoi and Cheng, Mengnan and Pan, Hailin and Zuo, Lulu and King, Samuel and others},
  journal={Cell},
  volume={187},
  number={26},
  pages={7351--7373},
  year={2024},
  publisher={Elsevier}
}

@article{staahl2016visualization,
  title={Visualization and analysis of gene expression in tissue sections by spatial transcriptomics},
  author={St{\aa}hl, Patrik L and Salm{\'e}n, Fredrik and Vickovic, Sanja and Lundmark, Anna and Navarro, Jos{\'e} Fern{\'a}ndez and Magnusson, Jens and Giacomello, Stefania and Asp, Michaela and Westholm, Jakub O and Huss, Mikael and others},
  journal={Science},
  volume={353},
  number={6294},
  pages={78--82},
  year={2016},
  publisher={American Association for the Advancement of Science}
}

@article{rodriques2019slide,
  title={Slide-seq: A scalable technology for measuring genome-wide expression at high spatial resolution},
  author={Rodriques, Samuel G and Stickels, Robert R and Goeva, Aleksandrina and Martin, Carly A and Murray, Evan and Vanderburg, Charles R and Welch, Joshua and Chen, Linlin M and Chen, Fei and Macosko, Evan Z},
  journal={Science},
  volume={363},
  number={6434},
  pages={1463--1467},
  year={2019},
  publisher={American Association for the Advancement of Science}
}

@article{wei2022single,
  title={Single-cell Stereo-seq reveals induced progenitor cells involved in axolotl brain regeneration},
  author={Wei, Xiaoyu and Fu, Sulei and Li, Hanbo and Liu, Yang and Wang, Shuai and Feng, Weimin and Yang, Yunzhi and Liu, Xiawei and Zeng, Yan-Yun and Cheng, Mengnan and others},
  journal={Science},
  volume={377},
  number={6610},
  pages={eabp9444},
  year={2022},
  publisher={American Association for the Advancement of Science}
}

@article{oliveira2024characterization,
  title={Characterization of immune cell populations in the tumor microenvironment of colorectal cancer using high definition spatial profiling},
  author={Oliveira, Michelli F and Romero, Juan P and Chung, Meii and Williams, Stephen and Gottscho, Andrew D and Gupta, Anushka and Pilipauskas, Susan E and Mohabbat, Syrus and Raman, Nandhini and Sukovich, David and others},
  journal={BioRxiv},
  pages={2024--06},
  year={2024},
  publisher={Cold Spring Harbor Laboratory}
}

@article{chen2022spatiotemporal,
  title={Spatiotemporal transcriptomic atlas of mouse organogenesis using DNA nanoball-patterned arrays},
  author={Chen, Ao and Liao, Sha and Cheng, Mengnan and Ma, Kailong and Wu, Liang and Lai, Yiwei and Qiu, Xiaojie and Yang, Jin and Xu, Jiangshan and Hao, Shijie and others},
  journal={Cell},
  volume={185},
  number={10},
  pages={1777--1792},
  year={2022},
  publisher={Elsevier}
}

@article{halmos2025dest,
  title={DeST-OT: Alignment of spatiotemporal transcriptomics data},
  author={Halmos, Peter and Liu, Xinhao and Gold, Julian and Chen, Feng and Ding, Li and Raphael, Benjamin J},
  journal={Cell systems},
  volume={16},
  number={2},
  year={2025},
  publisher={Elsevier}
}

@article{klein2023mapping,
  title={Mapping cells through time and space with moscot},
  author={Klein, Dominik and Palla, Giovanni and Lange, Marius and Klein, Michal and Piran, Zoe and Gander, Manuel and Meng-Papaxanthos, Laetitia and Sterr, Michael and Bastidas-Ponce, Aim{\'e}e and Tarquis-Medina, Marta and others},
  journal={BioRxiv},
  pages={2023--05},
  year={2023},
  publisher={Cold Spring Harbor Laboratory}
}

@article{peng2026stvcr,
  title={stVCR: spatiotemporal dynamics of single cells},
  author={Peng, Qiangwei and Zhou, Peijie and Li, Tiejun},
  journal={Nature Methods},
  volume={23},
  number={3},
  pages={542--553},
  year={2026},
  publisher={Nature Publishing Group US New York}
}

@article{huizing2026stories,
  title={STORIES: learning cell fate landscapes from spatial transcriptomics using optimal transport},
  author={Huizing, Geert-Jan and Samaran, Jules and Capocefalo, Daniele and Audit, Anna and Peyr{\'e}, Gabriel and Cantini, Laura},
  journal={Nature Methods},
  volume={23},
  number={3},
  pages={522--531},
  year={2026},
  publisher={Nature Publishing Group US New York}
}

@article{han20264d,
  title={4D single-cell spatial transcriptomics reveals dynamic morphogenetic gradients and regenerative domains in planarians},
  author={Han, Kai and Wang, Yuxiaofei and Li, Yao and Guo, Lidong and Chen, Yue and Liu, Xiawei and Lin, Yaru and Huang, Zhi and Liu, Qun and Guo, Wenjie and others},
  journal={GigaScience},
  volume={15},
  pages={giag064},
  year={2026},
  publisher={Oxford University Press}
}

@ArtifactSoftware{R,
    title = {R: A Language and Environment for Statistical Computing},
    author = {{R Core Team}},
    organization = {R Foundation for Statistical Computing},
    address = {Vienna, Austria},
    year = {2019},
    url = {https://www.R-project.org/},
}

\appendix

\section{Additional Implementation Details}

\subsection{Dataset Details}

We evaluate ST-FateFlow on three time-resolved spatial transcriptomic datasets covering distinct organisms, biological processes, temporal
scales, and spatial dimensions. A summary of the datasets and evaluation
time points is provided in Table~\ref{tab:dataset_details}.

\begin{table*}[t]
\centering
\caption{Summary of the datasets used in the interpolation experiments.
DPI denotes days post injury, hpa denotes hours post amputation, and dpa
denotes days post amputation.}
\label{tab:dataset_details}
\small
\setlength{\tabcolsep}{4.5pt}
\begin{tabular}{llllll}
\toprule
Dataset
& Organism
& Biological process
& Profiling platform
& Spatial dimension
& Evaluation time points \\
\midrule

ARTISTA
& Axolotl
& Telencephalon regeneration
& Stereo-seq
& 2D
& 5, 10, and 15 DPI \\

MOSTA
& Mouse
& Embryonic organogenesis
& Stereo-seq
& 2D
& E10.5 \\

PRISTA4D
& Planarian
& Whole-body regeneration
& Stereo-seq
& 3D
& 12 hpa, 36 hpa, 3, 5, 7, and 10 dpa \\

\bottomrule
\end{tabular}
\end{table*}

\paragraph{MOSTA}
The Mouse Organogenesis Spatiotemporal Transcriptomic Atlas
(MOSTA) was generated using Stereo-seq and contains single-cell-resolution
spatial transcriptomic profiles of sagittal mouse embryo sections from
embryonic day E9.5 to E16.5. The atlas
characterizes coordinated transcriptional and anatomical changes during
mouse organogenesis. We use the early organogenesis stages surrounding
E10.5 and treat E10.5 as an unobserved intermediate snapshot for
interpolation evaluation. The model predicts both the gene-expression
distribution and the two-dimensional organization of the embryo section
at E10.5.

\paragraph{ARTISTA}
The Axolotl Regenerative Telencephalon Interpretation via Spatiotemporal
Transcriptomic Atlas (ARTISTA) contains single-cell-resolution Stereo-seq
measurements of the axolotl telencephalon during brain regeneration. The regenerative series includes tissue sections
collected at multiple days post injury and captures substantial changes
in cell states and tissue organization following injury. We evaluate
interpolation at 5, 10, and 15 DPI. Each evaluation time point is removed
from the corresponding training sequence, and its observed cells are used
only as ground truth for evaluating the predicted gene-expression and
two-dimensional spatial distributions.

\paragraph{PRISTA4D}
The Planarian Regenerative Interactive Spatiotemporal Transcriptomic Atlas in four dimensions (PRISTA4D) provides single-cell-resolution spatial transcriptomic profiles of homeostatic and regenerating planarians. In addition to gene expression, PRISTA4D provides
reconstructed three-dimensional cell coordinates, allowing us to evaluate whether a model can recover continuous volumetric tissue remodeling. We
use snapshots from 0 hpa, 12 hpa, 36 hpa, 3 dpa, 5 dpa, 7 dpa, 10 dpa, and 14 dpa. The six internal stages---12 hpa, 36 hpa, 3 dpa, 5 dpa,
7 dpa, and 10 dpa---are evaluated as held-out interpolation targets. This dataset is particularly challenging because regeneration involves large-scale changes in both tissue geometry and cell-population composition.

\subsection{Preprocessing Details}

For each dataset, we first remove low-quality cells with insufficiently
detected genes and genes expressed in very few cells. Library sizes are
then normalized to $10{,}000$ counts per cell, followed by the
$\log(1+x)$ transformation. We identify the 2,000 most highly variable
genes while accounting for the temporal batches and retain the same gene
set across all time points within each dataset. This produces a common
gene-expression space in which cells from different snapshots can be
directly compared.

Spatial coordinates are centered and normalized within each dataset to
remove differences caused by coordinate scale. Two-dimensional
coordinates are retained for ARTISTA and MOSTA, whereas all three
coordinate dimensions are retained for PRISTA4D. ST-FateFlow directly
models the resulting 2,000-dimensional gene-expression vectors together
with the spatial coordinates. No PCA or other dimensionality reduction
is applied to the gene-expression input used by the velocity-field model.

For every interpolation experiment, the target snapshot is excluded from
model training. The model is conditioned on the observed snapshots
surrounding the target time and generates cells at the normalized
intermediate time. The generated cells are then compared with the
observed cells from the held-out snapshot. Because cells from different
experiments are inherently unpaired, all reported metrics compare
empirical cell distributions rather than assuming one-to-one cell
identities. The same preprocessing, temporal splits, and evaluation
samples are used for all applicable methods.

\section{Potential Analysis}

\subsection{Identification of Potential-Associated Genes}
\label{app:trend_score}

To identify genes whose expression changes systematically along the
learned potential landscape, we calculate a trend score that combines
the global expression--potential correlation with the expression
difference between high- and low-potential cells.

For gene $g$, the trend score is defined as
\begin{equation}
\operatorname{TrendScore}_g
=
r_g
\cdot
\operatorname{sign}(\Delta_g)
\cdot
\left|
\log_2 \operatorname{FC}_g
\right|,
\label{eq:trend_score}
\end{equation}
where
\begin{equation}
r_g
=
\operatorname{Pearson}
\left(
\mathbf{x}_g,\mathbf{P}
\right)
\label{eq:potential_correlation}
\end{equation}
is the Pearson correlation between the expression vector
$\mathbf{x}_g$ of gene $g$ and the learned cell potential
$\mathbf{P}$. The expression contrast between the high- and
low-potential groups is defined as
\begin{equation}
\Delta_g
=
\overline{x}_{g,\mathrm{high}}
-
\overline{x}_{g,\mathrm{low}},
\label{eq:potential_expression_difference}
\end{equation}
and the corresponding log fold change is
\begin{equation}
\log_2 \operatorname{FC}_g
=
\log_2
\left(
\frac{
\overline{x}_{g,\mathrm{high}}+\epsilon
}{
\overline{x}_{g,\mathrm{low}}+\epsilon
}
\right).
\label{eq:potential_fold_change}
\end{equation}
Here, $\overline{x}_{g,\mathrm{high}}$ and
$\overline{x}_{g,\mathrm{low}}$ denote the mean expression of gene $g$
among cells in the top and bottom potential quartiles, respectively.
We use the top $25\%$ of cells as the high-potential group and the
bottom $25\%$ as the low-potential group, with
$\epsilon=10^{-3}$ added for numerical stability.

The Pearson correlation measures whether gene expression changes
consistently across the complete potential range, whereas the absolute
log fold change measures the magnitude of the expression difference
between the two potential extremes. The factor
$\operatorname{sign}(\Delta_g)$ rewards agreement between the global
correlation and the high--low expression contrast. Consequently, a
large positive trend score indicates a strong and internally consistent
association with potential, regardless of whether the gene increases
or decreases with potential. Genes with inconsistent correlation and
high--low contrast receive negative scores.

The direction of association is determined separately using $r_g$:
genes with $r_g>0$ are classified as increasing with potential, whereas
genes with $r_g<0$ are classified as decreasing with potential. Within
each direction, genes are ranked according to
$\operatorname{TrendScore}_g$. This procedure is applied separately to
the cells associated with each examined differentiation path.

\begin{figure}[t]
    \centering


    \includegraphics[width=\columnwidth]
        {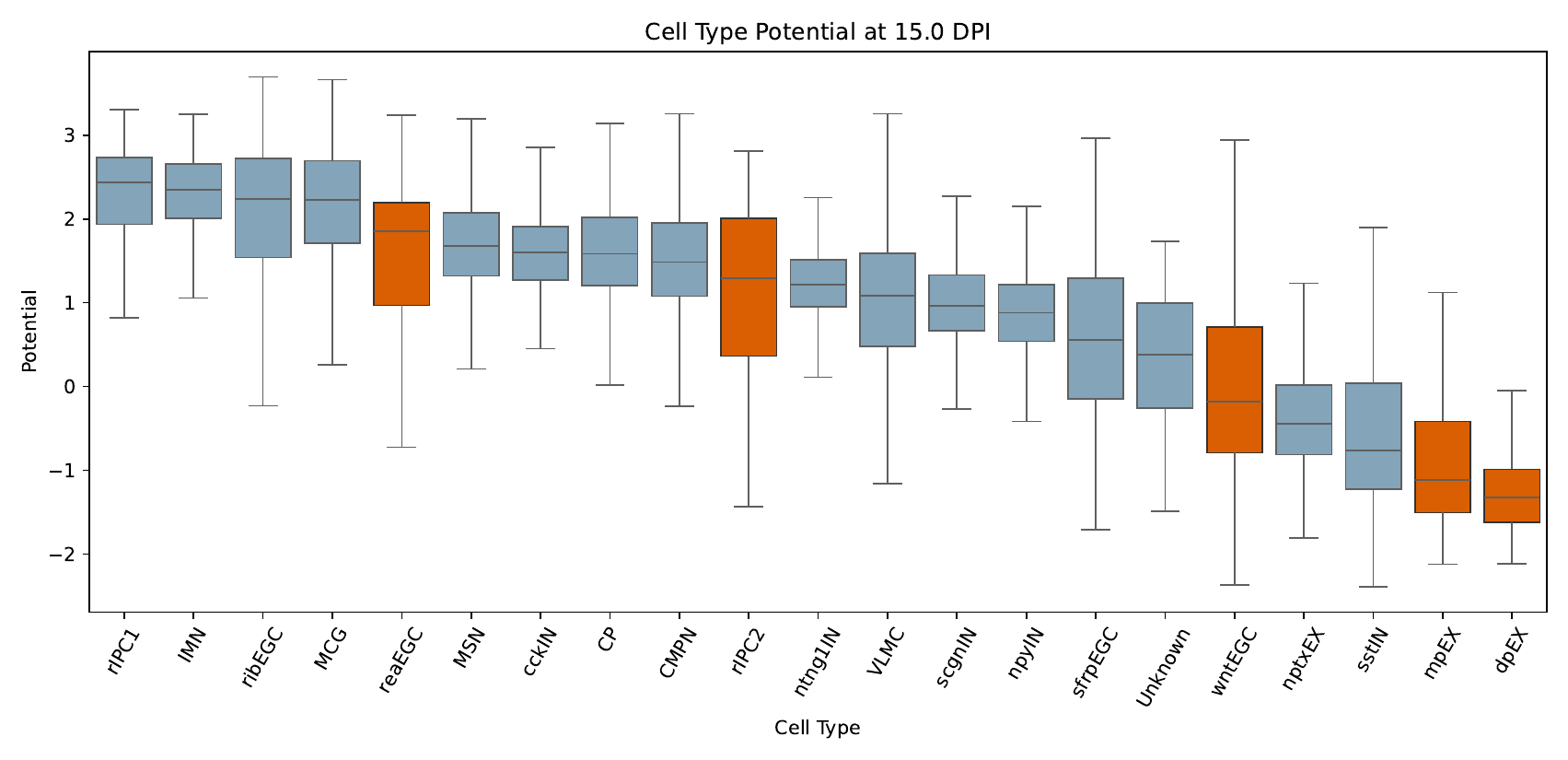}

    \caption{
        Cell-type potential distributions at 15.0 DPI. Cell types are ordered by their median potential values, red boxes denote the cell types of interest that exhibit clear differentiation trajectories
    }
    \label{fig:appendix_potential_boxplots}
\end{figure}

\end{document}